\documentclass[11pt,urlcolor=black, linkcolor=black]{article} 
\usepackage{cite}
\usepackage{amsmath, amsthm, amssymb,slashed, url}

\pdfoutput=1

\usepackage{ifpdf}
\ifpdf
  \usepackage[pdftex]{graphicx}
  \usepackage{epstopdf}
\else  
  \usepackage[dvips]{graphicx}
\fi

\usepackage[usenames,dvipsnames]{color}
\usepackage[colorlinks,citecolor=black]{hyperref}

\newcommand{\widesim}[2][1.5]{
  \mathrel{\overset{#2}{\scalebox{#1}[1]{$\sim$}}}
}
\parskip=\baselineskip
\def\Bbb{\mathbb}
\def\Tr{{\rm Tr}}
\def\16{{\bf 16}}

\def\1{{\bf 1}}
\def\2{{\bf 2}}
\def\3{{\bf 3}}
\def\4{{\bf 4}}

\def\x{{\sf x}}
\def\y{{\sf y}}

\def\bar{\overline}

\def\a{{\sf a}}

\def\tr{{\mathrm{tr}}}

\def\h{\widehat}
\def\bar{\overline}

\def\ra{\rangle}

\def\bp{\begin{pmatrix}}
\def\ep{\end{pmatrix}}
\def\la{\langle}
\def\R{{\Bbb{R}}}
\def\Z{{\Bbb{Z}}}

\def\S{{\mathcal S}}
\def\hat{\widehat}

\def\V{{\mathcal V}}
\numberwithin{equation}{section}

\def\d{\mathrm d}

\def\tT{{\widetilde T}}

\def\C{{\Bbb C}}
\def\Z{{\Bbb Z}}
\def\bar{\overline}
\def\X{{\mathcal X}}

\def\Res{{\mathrm{Res}}}
\def\bar{\overline}
\begin{document}

\def\be{\begin{equation}}
\def\ee{\end{equation}}
  
\def\aa{{\sf a}}
\def\bb{{\sf b}}
\def\a{{\bf a}}
\def\x{{\bf x}}
\def\y{{\bf y}}
\def\z{{\bf z}}
\def\d{{\mathrm d}}
 \def\sym{{\mathrm{sym}}}
 \def\zotimes{{\otimes N}}
 \def\cl{{\mathrm{cl}}}
\def\h{\widehat}
\def\t{\widetilde}
\def\CC{{\mathcal C}}
\def\tr{{\mathrm{tr}}}
\def\d{{\mathrm{d}}}
\def\H{{\mathcal H}}
\def\O{{\mathcal O}}
\def\Tr{{\mathrm{Tr}}}
\def\diag{{\mathrm{diag}}}
\def\vt{\vartheta}
\def\i{{\mathrm i}}
\def\la{\langle}
\def\ra{\rangle}
\def\ZZ{{\mathcal Z}}
\def\N{{\mathcal N}}
\def\Re{{\mathrm{Re}}}
\def\Im{{\mathrm{Im}}}
\def\S{{\mathcal S}}
\def\CC{{\Bbb C}}
\def\C{{\mathcal C}}
\def\VolH{V_H}
\def\SOS{{{\mathrm{SO}}(3)\times {\mathrm{SU}}(2)}}
\def\U{{\mathrm U}}
\def\Spin{\mathrm{Spin}}
\def\spinc{\mathrm{spin}_c}
\def\SU{\mathrm{SU(2)}}
\def\spinSU{\mathrm{spin \text{-} SU(2)}}
 \begin{titlepage}
\begin{flushright}
\end{flushright}
\vskip 1.5in
\begin{center}
{\bf\Large{Open Strings On The Rindler Horizon}}
\vskip
0.5cm { Edward Witten \\ \vskip.8cm {\small{ \textit{School of
Natural Sciences, Institute for Advanced Study,}\vskip -.4cm
{\textit{Einstein Drive, Princeton, NJ 08540 USA}}
}}}
\end{center}
\vskip 0.5in
\baselineskip 16pt
\abstract{It has been proposed that a certain $\Z_N$ orbifold, analytically continued in $N$, can be used to describe the thermodynamics of Rindler space in
string theory.    In this paper, we attempt to implement this idea for the open-string sector.   The most interesting result is that, although the orbifold is tachyonic for positive integer
$N$, the tachyon seems to disappear after analytic continuation to the region that is appropriate for computing $\Tr\,\rho^\N$, where $\rho$ is the density matrix of Rindler space
and $\Re\,\N>1$.   Analytic continuation of the full orbifold conformal field theory remains a challenge, but we find some evidence that if such analytic continuation is possible, the
resulting theory is a logarithmic conformal field theory, necessarily nonunitary.}
\date{January, 2018}
\end{titlepage}
\tableofcontents
\section{Introduction}\label{introduction}
In quantum field theory, the replica trick\footnote{This technique may have originated in spin glass theory.   See \cite{Parisi} for an introduction in that context.}
 is often used to compute entropies.  For example, to compute the entanglement entropy of quantum fields across  a Rindler horizon in $D$-dimensional
Minkowski spacetime,
one factors spacetime as $\R^2\times \R^{D-2}$ and replaces $\R^2$ by an $\N$-fold cover, for some integer $\N$,
branched over the origin in $\R^2$.   A formal argument shows that a path integral on this cover computes
$\Tr\,\rho^\N$, where $\rho$ is the density matrix for an observer outside the horizon.

To compute the Rindler entropy, one would like to analytically continue $f(\N)=\Tr\,\rho^\N$ as a function of $\N$ to the vicinity of $\N=1$, and then differentiate with respect to $\N$ to get the entropy
$-\Tr\,\rho\log\rho=-f'(1)$.  In this particular example, one can directly access non-integer $\N$ by considering a cone with a general opening angle.   However, let us continue the discussion
assuming that this option is not available, as that will be the case in our problem of interest.
Then one reasons as follows.  If $\rho$ is a density matrix, that is, a positive hermitian matrix satisfying $\Tr\,\rho=1$,
then $f(\N)=\Tr\,\rho^\N$ is holomorphic in $\N$ for $\mathrm{Re}\,\N\geq 1$, and bounded in absolute value by 1 in that half-plane.  
Therefore, we expect that the function $f(\N)$, initially defined for integer $\N$,
will have an analytic continuation to the half-plane $\mathrm{Re}\,\N\geq 1$, bounded by 1.   By Carlson's theorem,\footnote{The
theorem says that a function $f(N)$ that is holomorphic for $\Re\,N\geq 1$, vanishes for positive integer $N$, and obeys certain bounds on its growth is identically zero.
The required bounds are that $f(N)$ is exponentially bounded throughout the half-plane $\mathrm{Re}\,N\geq 1$, meaning that $|f(N)|<C\exp(\lambda|N|)$ for some constants $C,\lambda$,
and moreover that on the line $\Re\,N=1$, such a bound holds with $\lambda<\pi$.
 See \cite{Boas}, p. 153.   The growth conditions exclude  counterexamples such as $\sin\,\pi N$, $\frac{e^{-N}}{N}\sin\,\pi N$ and $e^{N^2} \sin\pi N$.  Carlson's theorem
has an obvious analog for a function that vanishes on the positive odd integers, which can be reduced to the case of a function that vanishes on all positive integers by the mapping
$N\to 1+(N-1)/2$. It is this version of the theorem that we will actually use later, since we will consider a $\Z_N$ orbifold for odd $N$.} such a continuation
is unique if it exists, so once we know $\Tr\,\rho^\N$ for positive integer $\N$, we can determine it throughout the half-plane $\Re\,\N\geq 1$ by finding the appropriate analytic continuation.   Then
taking a derivative at $\N=1$ will give the entropy.

In local quantum field theory in a background spacetime, this procedure is not rigorously valid because  observations outside the horizon cannot really be described
by a density matrix (basically because
the algebra of observables outside the horizon is a von Neumann algebra of Type III, not of Type I).
Still, in practice, this procedure frequently gives good results.   

In string theory, we cannot apply the replica trick in precisely the fashion stated, because there is no known conformal field theory whose target space is an $\N$-fold cover of $\R^2$, branched
over the origin.   However, an alternative has been suggested in the form of an $\R^2/\Z_N$ orbifold \cite{Dabholkar},
where, for technical reasons, we will assume $N$ to be odd; the even $N$ case is slightly different.\footnote{For a previous attempt to use
this orbifold to compute the Rindler entropy, see \cite{Takayanagi}.  For alternative approaches to entanglement entropy in string theory, see \cite{Uglum, Hartnoll, Donnelly, Bala}.}

  Now, formally, the orbifold computes $\Tr\,\rho^{1/N}$.    Again, one might hope to
analytically continue from odd integer values of $N$ and eventually to compute the entropy by differentiating at $N=1$.  
This program runs into a variety of difficulties.     Formally, $\Tr\,\rho^{1/N}$ should be given by the exponential of the sum of amplitudes for connected string worldsheets, so the (unexponentiated) sum of connected
worldsheets should compute $\log\,\Tr\,\rho^{1/N}$.   The 1-loop approximation $\left.\log \Tr\,\rho^{1/N}\right|_1$ would be
\be\label{zoffo}\left.\log\,\Tr\,\rho^{1/N}\right|_1= \ZZ_{1,N}, \ee
where $\ZZ_{1,N}$ is the 1-loop vacuum amplitude of the orbifold.
 
However, the closed-string spectrum of the orbifold is tachyonic, and as a result the integral in (\ref{zoffo})
is infrared-divergent. This statement holds for both closed and open strings, though there is no tachyon in the open-string spectrum. In an open-string 1-loop amplitude,
 the divergence results from  closed-string exchange in the
crossed channel.

This problem in computing $\Tr\,\rho^{1/N}$ for $N>1$
 should not be too discouraging, because in quantum mechanics in general, $\Tr\,\rho^{1/N}$ can be divergent if $N>1$.  
But it does mean that we  need a more refined starting point for analytic continuation.   For this, as in \cite{Dabholkar, Takayanagi}, we use the
standard representation of  $\ZZ_{1,N}$ as an integral over a modular parameter.   For open strings, the formula reads
\be\label{woffol} \ZZ_{1,N}=\frac{1}{2}\int_0^\infty\frac{ \d T}{T} \,Z_N(T),\ee
where $T$ is a real proper time parameter and $Z_N(T)$ is the 1-loop partition function with modular parameter $T$.   (To lighten notation, we write this as $Z_N(T)$ rather than $Z_{1,N}(T)$.  Similarly, we will sometimes write, for example, just $\Tr\,\rho^{1/N}$, not $\left.\Tr\,\rho^{1/N}\right|_1$.)
This formula has a well-known analog for closed strings.     

Moreover, one can write an explicit formula for $Z_N(T)$ or its closed-string analog.    The question then arises of whether one can analytically continue these functions away from positive odd 
integer values of $N$ in a reasonable way.  The most interesting version of the problem is for closed strings, since the graviton is described by a closed string.
But that version of the problem appears more difficult and in this paper we will consider open strings.  
In fact, we will consider the simplest open-string problem, which is the case of a Dp-brane whose worldvolume crosses the Rindler horizon.
Thus in Type II superstring theory on $\R^2/\Z_N\times \R^8$, we consider a Dp-brane with worldvolume $\R^2/\Z_N\times \R^{p-1}$.

In this context, it is possible to explicitly describe the unique continuation of $Z_N(T)$ to the half-plane $\Re\,N\geq 1$, with bounded growth, 
that is allowed by Carlsen's theorem.  As a bonus, it turns out that this analytically-continued function is actually holomorphic
in a larger half-plane $\Re\,N>0$. Note that $\Re\,N>0$ is equivalent to $\Re\,\N>0$ where $\N=1/N$.   
So in particular, we can analytically continue to $\Re\,\N>1$.     That is the region where we really
hope for good behavior, since $\Tr\,\rho^{1/N}=\Tr\,\rho^\N$, and in quantum mechanics in general, $\Tr\,\rho^{\N}$ is holomorphic for $\Re\,\N>1$.   

As usual, one can extract some information about the closed-string sector by studying the open-string partition function $Z_N(T)$ for small $T$.
It turns out that to the extent that the closed-string sector can be probed in this way, it is tachyon-free for $\Re\,\N>1$.  This means that there
is no exponential divergence in the integral that is supposed to compute $\log \,\Tr\,\rho^{\N}$.  The integral may still have a power-law divergence for $T\to 0$ due to the exchange of massless closed-string states. As always in string theory, such a divergence is an infrared divergence.   
In fact, we should expect such a divergence if $p$ is too large.
Recall that 1-loop scattering amplitudes of modes that propagate on the Dp-brane worldvolume are infrared-divergent for $p>6$, because 
the Dp-brane acts as a source for massless bosons (the dilaton and graviton) that propagate in bulk, and this produces infrared-divergent effects if the number of
dimensions transverse to the Dp-brane is too small.   It turns out, however, that the bound we need to make $\Tr\,\rho^{\N}$ infrared-finite is stronger than
$p\leq 6$.  For $\Re\,\N>1$, $\Tr\,\rho^\N$ turns out to be infrared-finite for $p\leq 4$.   
After differentiating with respect to $\N$ and setting $\N=1$, one finds
that $\Tr\,\rho\log\rho$ is infrared-finite for $p\leq 2$.   The proper interpretation  is far from clear.
However, at a technical level, it appears that the unexpected behavior may be related to the fact
that  the orbifold
theory, analytically continued away from positive  integer $N$, is a nonunitary logarithmic conformal field theory (see \cite{RS,G} for early
work, and \cite{C} for a recent review).  As we will explain,  a logarithmic conformal field theory
 can potentially lead to the sort of behavior that we will actually find.

In section \ref{strategy}, we describe our strategy for analytic continuation.    Section \ref{partfn} consists of a derivation of standard formulas
for the Dp-brane partition function.   In section \ref{comprho}, we combine these ingredients to compute $\Tr\,\rho^\N$ and to investigate
some of its properties.  The paper also contains two appendices.    Appendix \ref{reform} describes a possible starting point for analytic continuation.
Appendix \ref{pfunction} describes an alternative formula for the partition function. 

The results in this paper will probably only be useful if it turns out that the closed-string version of the problem has a reasonable analytic continuation.
If that is the case, and the analytically continued closed-string theory is indeed tachyon-free for $\Re\,\N>1$, then presumably the one-loop closed-string
contribution to $\Tr\,\rho^\N$ will be well-defined and finite.   Even so, there would be a potential obstacle in going to higher orders: it will be possible to get
sensible answers for multiloop contributions to $\Tr\,\rho^\N$ only if dilaton tadpoles vanish.\footnote{A possible reason for such vanishing is as follows.
A tadpole arises when a Riemann surface $\Sigma$ degenerates to two components $\Sigma_\ell$ and $\Sigma_r$ glued at a point $p$.   Writing $\V$ for the dilaton
vertex operator, the tadpole amplitude is proportional to a product $\langle \V(p)\rangle_{\Sigma_\ell} \langle \V(p)\rangle_{\Sigma_r}$.   At $N=1$, both factors
vanish due to supersymmetry.   Differentiating with respect to $N$ and then setting $N=1$ will potentially eliminate the vanishing of one of the two factors but not both.
This suggests that a dilaton tadpole may not present a problem in multi-loop contributions to the entropy.   However, the same reasoning suggests that there is such a difficulty
in multi-loop contributions to $\Tr\,\rho^\N$ with $\Re\,\N>1$.   The potential problem is an infrared divergence due to gravitational and dilaton back-reaction that is sourced by 
the tension of the Rindler horizon; this tension vanishes at tree-level
assuming that the orbifold CFT can be analytically continued in $N$, but it may be nonzero (for $N\not=1$) in higher orders.}
Note also that we compute the annulus contribution to $\Tr\,\rho^N$ without first discussing the disc contribution.   The reason for this is simply that a proper
framework for the disc contribution is not clear.   For a comment on this contribution, see the conclusion of section \ref{comprho}.

\section{Strategy For Analytic Continuation}\label{strategy}

The 1-loop amplitude for open strings is computed using a worldsheet that is an annulus $I\times S^1$, where $I$ is an interval and $S^1$ a circle.
The annulus partition function for strings in $\R^{10}$ can be written as a trace in the open-string Hilbert space (but vanishes because of spacetime supersymmetry).
The annulus partition function for strings on the orbifold is obtained in a standard way as a sum of traces.   Let $U$ be a generator of the orbifolding group $\Z_N$
and, for integer $k$,  let $Z_{k,N}$ be a trace in the open string Hilbert space with an insertion of $U^k$ (see section \ref{partfn} for more detail).    
The orbifold partition function on the annulus is then
\be\label{whaft} Z_N(\tau)=\frac{1}{N}\sum_{k=0}^{N-1}Z_{k,N}(\tau), \ee
where $\tau$ is the modulus of the annulus.
The factor of $1/N$ reflects the fact that the projection operator onto $\Z_N$-invariant states is $\frac{1}{N}\sum_{k=0}^{N-1} U^k$.

For analytic continuation, we will use the fact that there is  a meromorphic function $J(z,\tau)$ with the following properties:
\begin{enumerate}\label{zom}
\item{The twisted partition function $Z_{k,N}(\tau)$ can be expressed in terms of $J$ by
\be\label{zolb} Z_{k,N}(\tau)=J(k/N,\tau).\ee}
\item{$J(z,\tau)$ is a periodic and even function 
\be\label{olb} J(z+1,\tau)=J(z,\tau)=J(-z,\tau). \ee}
\item{The residues at poles of $J(z,\tau)$ vanish exponentially for $\Im\,z\to\pm \i\infty$.   (More precisely, $J$ and its residues behave near $\pm\i\infty$ in a way that
permits the following computation.) Moreover, $J(0,\tau)=0$, because
of spacetime supersymmetry.}
\item{Poles of $J(z,\tau)$ are at $\Re\,z=0$ or $1/2$ mod $\Z$.   They are all simple poles except for a double pole at $z=1/2$.}
\end{enumerate}
The existence of such a  $J$ is a restatement of standard facts, and will be explained in section \ref{partfn}.

The orbifold partition function on the annulus is therefore
\be\label{wolb} Z_N(\tau) =\frac{1}{N}\sum_{k=1}^{N-1} J(k/N,\tau), \ee
where we omit the term with $k=0$ since $J(0)=0$.   

Consider the function
\be\label{plob} K(z,N)=\sum_{k=1}^{N-1}\frac{\pi \sin\pi z}{\sin (\pi k/N)\sin\pi(z-k/N)}.\ee
It is a periodic function, $K(z+1,N)=K(z,N)$, and  bounded for $\Im\,z\to\pm \infty$.  
The poles of $K(z,N)$ in the strip $0\leq \Re\,z\leq 1$ are simple poles of residue 1
at $z=k/N$, $k=1,\cdots,N-1$.

Now let us make the periodic identification $z\cong z+1$ to define a cylinder, and view the function $K(z,N)J(z,\tau)$
as a meromorphic function on the cylinder.  
We can view the orbifold partition function on the annulus as a sum of residues of the function $KJ$ at the poles of $K$
on the cylinder:
\be\label{wilob} Z_N(\tau) =\frac{1}{N}\sum_{k=1}^{N-1}\mathrm{Res}_{z=k/N}\,( K(z,N)J(z,\tau)). \ee
On the other hand, the sum over all residues of this function on the cylinder vanish.  Here  we need a condition on the
behavior of $J(z,\tau)$ for $z\to \pm \i\infty$; the formulas in section \ref{partfn} will make it clear that there is no problem.
Therefore, we can get an alternative formula for $Z_N(\tau)$ as minus the sum of residues at poles of $J$.  (Since the poles
of $J$ are at $\Re\,z=0,1/2$, the set $\S$ of poles of $J$ on the cylinder is disjoint from the set $\{1/N,2/N,\cdots,(N-1)/N\}$
of poles of $K$.) Thus
\be\label{nilob} Z_N(\tau)=-\frac{1}{N}\sum_{z_0\in\S}\Res_{z=z_0}\bigl(K(z,N)J(z,\tau)\bigr). \ee

If the poles of $J(z,\tau)$ were all simple poles, the formula would reduce to $-\frac{1}{N}\sum_{z_0\in\S}  K(z_0,N)\Res_{z=z_0}J(z,\tau)$.
Actually, $J(z,\tau)$ has a double pole at $z=1/2$, so setting $\S'$ to be the set of simple poles of $J(z,\tau)$, the correct version of the formula is
\begin{align}\label{munno} Z_N(\tau)=&-\frac{1}{N}\sum_{z_0\in\S'}K(z_0,N)\Res_{z=z_0}J(z,\tau)\cr &-\frac{1}{N} \left.\partial_z K(z,N)\right|_{z=1/2}\Res_{z=1/2}\left((z-1/2) J(z,\tau)\right). \end{align}
(This treatment of the double pole at $z=1/2$ uses the fact that $J(z-1/2)=J(-(z-1/2))$, by virtue of eqn. (\ref{olb}), so near $z=1/2$,
$J(z,N)$ behaves as $\frac{c}{(z-1/2)^2}+\dots$, where $c$ is a constant and the omitted terms are regular.)

From these formulas, it is clear that to analytically continue $Z_N(\tau)$, all we need is an analytic continuation of $K(z,N)$.   By Carlson's theorem, if there is an analytic continuation of
$K(z,N)$ that is holomorphic for $\Re\,N\geq 1$ and obeys suitable exponential bounds, then this analytic continuation is unique.   In trying to find this analytic
continuation, one runs into a slight surprise.  For fixed $z$, provided $\Re\,z$ is suitably constrained, the analytic continuation in $N$ suggested by Carlson's theorem does exist
and can be written explicitly.    But one has to use different continuations for different values of $\Re\,z$.  There is not a holomorphic function of $z$ and $N$ that coincides with $K(z,N)$
when $N$ is an integer and satisfies the conditions of Carlson's theorem as a function of $N$ for fixed $z$.   

The poles of $J(z,\tau)$ are all at $\Re\,z=0$ or $\Re\,z=1/2$, so we only need to analytically continue $K(z,N)$ at those values of $\Re\,z$.   At $\Re\,z=0$, we proceed as follows.
First, for integer $N$, an alternative formula for $K(z,N)$ is
\be\label{wombo} K_1(z,N)=\pi N\cot\pi Nz -\pi \cot \pi z. \ee
For $\Re\,z=0$, $K_1(z,N)$ is holomorphic in $N$ for $\Re\,N\geq 1$, and satisfies the appropriate exponential bounds.   It is the unique continuation of $K(z,N)$
from positive integer values of $N$ that has those properties.    As a bonus, for $\Re\,z=0$,  $K_1(z,N)$ is actually holomorphic in a larger half-plane $\Re\,N>0$.   

For $\Re\,z=1/2$, we have to proceed more carefully and consider analytic continuation from {\it odd} positive integer values of $N$. (This is natural because we will study an orbifold defined
for odd $N$; an analogous orbifold for even $N$ is slightly different \cite{Dabholkar}.)  For $N$ an odd positive integer, an alternative formula
for $K(z,N)$ is
\be\label{tombo} K_2(z,N)=\pi N\cot\left(\pi(N(z-1/2)+1/2) \right)-\pi\cot\pi z. \ee
For $\Re\,z=1/2$, $K_2(z,N)$ is the unique continuation of $K(z,N)$ from positive odd integer values of $N$ that is holomorphic in the half-plane $\Re\,N>1$ and satisfies the appropriate
exponential bounds.   Again, as a bonus, for $\Re\,z=1/2$, $K_2(z,N)$ is holomorphic in the larger half-plane $\Re\,N>0$.

We can now write a version of  eqn. (\ref{munno}) that exhibits analytic continuation in $N$.
If $\S_0$ is the set of poles of $J(z,\tau)$ with $\Re\,z=0$, $\S_{1/2}$ is the set of poles with $\Re\,z=1/2$, and $\S'_{1/2}$ is the subset of $\S_{1/2}$ with $z\not=1/2$, 
then an analytic continuation of $Z_N(\tau)$ that is holomorphic for $\Re\,N>0$ is
\begin{align}\label{zunno} Z_N(\tau)=&-\frac{1}{N}\sum_{z_0\in\S_0}K_1(z_0,N)\Res_{z=z_0}J(z,\tau)-\frac{1}{N}\sum_{z_0\in\S_{1/2}'} K_2(z_0,N)\Res_{z=z_0} J(z,\tau)\cr &- \frac{1}{N}\left.\partial_z K_2(z,N)\right|_{z=1/2}\Res_{z=1/2}\left((z-1/2) J(z,\tau)\right). \end{align}

As is obvious in the formulas we have written, $K_1(z,N)$ and $K_2(z,N)$ can be continued away from the lines $\Re\,z=0$ and $\Re\,z=1/2$ to get meromorphic functions of $z$ and $N$.
But those functions are different.  In this sense, trying to analytically continue $K(z,N)$ in the fashion suggested by Carlson's theorem leads to 
different  results depending on the assumed value of $\Re\,z$. 

More generally, if one wants an analytic continuation of $K(z,N)$ along the line $\Re\,z=a/b$, where $a,b$ are relatively prime integers, one can restrict to $N$ congruent to $b$ mod 1 and
use the function
 \be\label{bulo}K_{[a/b]}(z,N)=N\cot\left(\pi(N(z-a/b)+a/b)\right)-\cot \pi z. \ee
 This has similar properties to $K_1(z,N)$ and $K_2(z,N)$.  

  In section \ref{comprho}, we will learn that to the extent that the closed-string sector can be probed
from a knowledge of the open-string partition function, it is tachyon-free if $K_2(z,N)$, after analytic continuation in $z$, is free of poles in the strip $0<\Re\,z<1$. From
the formula in (\ref{tombo}), we see that to get a pole in $K_2(z,N)$, we need $N(z-1/2)=k-1/2$ for some integer $k$.   Equivalently, with $\N=1/N$, we need
\be\label{holfe}\N=\frac{z-1/2}{k-1/2}. \ee
For $k\in\Z$ and $0<\Re\, z<1$, the right hand side has real part less than 1, so there is no pole in the strip if  $\Re\,\N\geq 1$.   This is why, as stated in the introduction,
evaluation of the 1-loop open-string contribution to  $\Tr\,\rho^\N$ for $\Re\,\N\geq 1$ does not run into a closed-string tachyon.

Because $J(0,\tau)=0$, we would still get a correct formula for $Z_N(\tau)$ if we add a multiple of $\cot\pi z$ to the original function $K(z,N)$; this
would shift the coefficient of the $\cot\pi z$ terms in $K_1(z,N)$ and $K_2(z,N)$ by an equal amount.   We have chosen this coefficient to try to make the formulas of section \ref{comprho} as natural-looking
as possible.  To be more precise, the coefficient was chosen so that an indication of non-diagonalizability of the closed-string spectrum in the $K_1$ contribution (which we tentatively
identify with the Ramond-Ramond sector of closed strings) only occurs for massive
states.   It is conceivable that this was not the best choice.

\section{Dp-Brane Partition Function}\label{partfn}

\begin{figure}
 \begin{center}
   \includegraphics[width=3.5in]{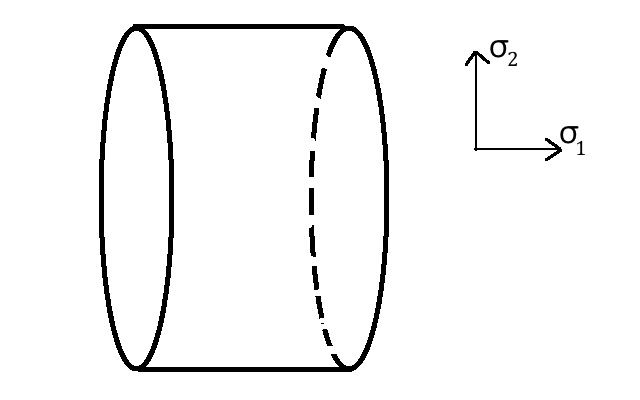}
 \end{center}
\caption{\small An annulus  in which $\sigma_1$  parametrizes an interval and $\sigma_2$ parametrizes a circle describes an open string propagating in the $\sigma_2$ direction, or a closed string propagating in the $\sigma_1$ direction,
known as the crossed channel.   The closed string states that contribute all have $L_0=\bar L_0$, since the annulus is invariant under rotation of the circle parametrized by $\sigma_2$. \label{Annulus}}
\end{figure}

In this section, we will study open strings on an annulus $I\times S^1$, where $I$ is the interval $0\leq \sigma_1\leq \pi$ and $S^1$ is the circle $\sigma_2\cong \sigma_2+2\pi T$ (fig. \ref{Annulus}).
For the metric of the annulus, we take $\d s^2=\d\sigma_1^2+\d\sigma_2^2$.   A path integral on the annulus has two useful interpretations.    It describes an open string, of width $\pi$, parametrized
by $\sigma_1$,
propagating in the $\sigma_2$ direction for an imaginary  proper time $2\pi T$.    The initial and final states of the open string are identified to make a trace.  Alternatively, the same annulus describes a closed
string, of circumference $2\pi T$, parametrized by $\sigma_2$, propagating in the $\sigma_1$ direction for an imaginary proper time $\pi$.   In this description, there is no trace; the closed
string is created from a ``brane'' at $\sigma_1$ =0 and annihilated on the brane at $\sigma_1=\pi$.
However, using conformal invariance,
it is convenient to rescale lengths by a factor $1/T$
so that the closed string has a standard circumference $2\pi$.    Then  $\sigma_1$, which is the proper time in the closed-string picture, has a range 
$\pi/T$.   This is  $2\pi \tT$ where $\tT$ is the closed-string
proper time parameter
\be\label{zofox}\tT=\frac{1}{2T}.\ee
The small $T$ behavior of the annulus -- which corresponds to the ultraviolet region in field theory -- is best understood in terms of the large $\t T$ behavior of the closed-string description.

In section \ref{First}, we construct the annulus partition function for the Dp-brane.   In section \ref{Second}, we compute the residues that are needed to evaluate the formula (\ref{zunno}) for the
analytically continued partition function. In section \ref{closed}, we discuss the closed-string sector.

\subsection{Computing the Partition Function}\label{First}

We will compute the annulus partition function for a Dp-brane in the  orbifold using light-cone Green-Schwarz fermions.\footnote{What follows
 is a standard type of computation which we present for completeness;
an equivalent set of formulas for this particular orbifold has been described and discussed in \cite{Dabholkar,Takayanagi}.}
These are a set of eight world-sheet fermions that transform as a spinor of a copy of $\Spin(8)$ that acts on, say, the first eight coordinates in $\R^{10}$.  
We then construct an $\R^2/\Z_N\times \R^8$ orbifold in which $\Z_N$ acts on, say, the first two coordinates, which we will call $X_1$ and $X_2$.  Consider
an element of $\Z_N$ that rotates $\X=X_1+\i X_2$ by a phase $\exp(2\pi \i r/N)$, for some integer $r$.    The Green-Schwarz fermions will then consist of four
complex fermions that are rotated by a phase that is the square root of this.   Here it is important whether $N$ is odd or even.   If $N$ is odd, then a square root
of $\exp(2\pi\i r/N)$ is again an $N^{th}$ root of 1, which one can write as $\exp(2\pi\i r(1-N)/2N)$.   We will focus on this case, both because it is slightly simpler and because
we are interested in analytically continuing to $N=1$, which is odd, so it is more natural to start with odd values of $N$.  (See \cite{Dabholkar} for a description of the corresponding orbifolds
with even $N$.)

It is convenient to set $k=r(1-N)/2$, which is an integer assuming that $N$ is odd.   Then a general element of  $\Z_N$ acts on the four complex fermions by a phase $\exp(2\pi \i k/N)$, and on
$\X$ by a phase $\exp(4\pi\i k/N)$.    The $\Z_N$ element with $k=1$ mod $N$ is a generator $U$ of the $\Z_N$ group.

Let $\H$ be the Hilbert space of open strings on $\R^{10}$, and $H$ the corresponding Hamiltonian.   The annulus partition function on $\R^{10}$ in the Green-Schwarz description is
\be\label{moffo} \Tr_\H\,(-1)^F\exp(-2\pi H T) . \ee
This vanishes because of supersymmetry.
The open-string Hilbert space $\H_N$ of the orbifold is just the $\Z_N$-invariant subspace of $\H$.   The projector onto the $\Z_N$-invariant subspace is $P=\frac{1}{N}\sum_{k=0}^{N-1}U^k$,
so the annulus partition function is
\begin{align}\label{woffo}Z_N=&\Tr_{\H_N} \,(-1)^F\exp(-2\pi HT) =\Tr_\H\,(-1)^F P\exp(-2\pi HT)\cr  =&\frac{1}{N}\sum_{k=0}^{N-1}\Tr_\H \,U^k (-1)^F\exp(-2\pi HT) \end{align}
The $k=0$ term in this sum coincides with (\ref{moffo}), up to a factor of $1/N$, and so vanishes.    Thus
\be\label{offo}Z_N=\sum_{k=1}^{N-1}Z_{k,N},~~~~~Z_{k,N}= \frac{1}{N}\Tr_\H\, U^k(-1)^F\exp(-2\pi HT). \ee

We want to compute the annulus partition function for open strings that end on a Dp-brane whose world-volume is spanned by $X_1,\cdots, X_{p+1}$.   Except for bosonic
zero-modes, the value of $p$ is not important.  One way to see this is to observe that if one toroidally compactifies the dimensions $X_{p+2},\cdots ,X_{10}$, then $T$-duality
on some of those coordinates will change the value of $p$ without affecting the spectrum of worldsheet modes other than the bosonic zero-modes.   For the bosonic zero-modes,
there definitely is a dependence on $p$.

As usual, modes of a free field on an interval   of width $\pi$  are conveniently compared to modes of a chiral free field on a circle of circumference $2\pi$  
that is made by taking two copies of $I$ glued together at the endpoints.   This circle can be parametrized by an angular variable $\sigma_1$, $0\leq \sigma_1\leq 2\pi$.
  Together $\sigma_1$ and $\sigma_2$ parametrize a torus $\Sigma$ with a modular parameter $\tau=\i T$ and therefore\footnote{Thus 
$q$ is positive; in what follows, $q^\alpha$, for a real number $\alpha$, will represent the positive number $\exp(-2\pi\alpha T)$.}
$q=\exp(2\pi\i \tau)=\exp(-2\pi T)$.  In computing  $Z_{k,N}$ from eqn. (\ref{offo}),  all bosons and fermions obey periodic boundary conditions
in the $\sigma_1$ and $\sigma_2$ directions, except for the twist in the $\sigma_2$ direction associated to the orbifolding.

The twisted partition function $Z_{k,N}$ is a product of factors coming from bosons and fermions:
\be\label{uvv}Z_{k,N}=Z_{k,N}^F Z_{k,N}^B.\ee
First let us compute the fermionic partition function $Z_{k,N}^F$.
 Let $\psi,\t\psi$ be a conjugate pair of chiral fermions, twisted by angles $\pm 2\pi k/N$.
For every positive integer $n$, the mode expansion of either  $\psi$ or $\t\psi$ contains a creation operator that raises the energy by $n$ units.   The contribution of these
modes to the partition function is a factor
\be\label{tofo} (1-q^n\exp(2\pi\i k/N)) (1-q^n\exp(-2\pi \i k/n)). \ee
More subtle is the role of the fermion zero-modes.    Both $\psi$ and $\t\psi$ have a zero-mode, transforming respectively as $\exp(\pm 2\pi\i k/N)$ under the twist.
Quantizing these two modes gives a pair of states, of opposite statistics, transforming under the twist as $\exp(\pm \pi\i k/N)$.   The contribution to the partition function of these
modes is thus a factor 
\be\label{lofo} \pm(\exp(\pi \i k/N) -\exp(-\pi\i k/N))=\pm 2\i \sin(\pi k/N), \ee
where the overall sign depends on which of the two states is bosonic and which is fermionic.  Out of context, there is no natural choice of this sign, a fact that is related to the fact
that $\exp(\pi\i k/N)$ is a square root of the twisting phase $\exp(2\pi\i k/N)$, but there is no natural way to choose the sign of the square root.  This ambiguity will disappear in a moment
when we combine the contributions of the four complex fermions.

Finally, we should remember that the ground state energy of a  complex chiral fermion that is periodic on the circle is $+1/12$.  Altogether, the partition function of
such a complex fermion is
\be\label{nofo} \pm 2\i \sin(\pi k/N) q^{1/12}\prod_{n=1}^\infty  (1-q^n\exp(2\pi\i k/N) )(1-q^n\exp(-2\pi \i k/n)).\ee
Combining four such multiplets, we get the fermionic partition function 
\be\label{zoofo}Z^F_{k,N}=  (2\sin (\pi k/N))^4 q^{1/3}\prod_{n=1}^\infty   (1-q^n\exp(2\pi\i k/N))^4 (1-q^n\exp(-2\pi \i k/n))^4.\ee
Here the sign is correct.   This may be seen as follows.   When we expand $(2\sin(\pi k/N))^4=(2\i (e^{\i \pi k/N}-e^{-\i \pi k/N}))^4$ in powers of $\exp(2\pi\i k/N)$, only integer powers appear,
so we do not need to choose a square root.  The maximum power that appears is $\exp(2\pi \i k/N)^2=\exp(4\pi ik/N)$.   This is the same as the phase by which the given $\Z_N$ element
acts on $\X=X_1+\i X_2$, so it corresponds to angular momentum 1 in rotation of that plane.   The corresponding spin 1 mode is a gauge boson mode, which is bosonic,
 so it should appear with a positive coefficient.  We have chosen the sign in eqn. (\ref{zoofo}) to ensure this.

We can conveniently write
\be\label{loofo}Z^F_{k,N}= F^4(z,\tau),\ee
with
\be\label{noofo} z=k/N\ee
and
\be\label{nuco}F(z,\tau) =  (w^{1/2}-w^{-1/2}) q^{1/12}\prod_{n=1}^\infty  (1-q^n w) (1-q^n w^{-1}),~~~w=\exp(2\pi\i z).\ee

Now let us consider $Z_{k,N}^B$.   The nonzero modes of the bosons
can be treated rather as before.   First consider the complex boson field $\X$ that is twisted by $\exp(4\pi\i k/N)$.      For every positive integer $n$,
both $\X$ and $\bar \X$ had a mode of energy $n$.  The partition function of these modes is
\be\label{dofo}\frac{1}{1-q^n\exp(4\pi\i k/N)}\frac{1}{1-q^n \exp(-4\pi\i k/N)}. \ee
This is just the inverse of the corresponding factor for fermions, except that the twist angle is twice as large.   
Similarly, the ground state energy of a complex boson (untwisted in the $\sigma_1$
direction) is $-1/12$, which will give a factor $q^{-1/12}$.    We also have to include in light cone gauge the nonzero modes of six untwisted bosons $X_3,\cdots ,X_8$.
Including the ground state energy, these give a factor $1/\eta^6(\tau)$, where $\eta$ is the Dedekind eta function $\eta(\tau)=q^{1/24}\prod_{n=1}^\infty(1-q^n)$.
Altogether, then the nonzero modes of the bosons give
\be\label{wolbo}\frac{1 }{q^{1/12}\prod_{n=1}^\infty\left((1-q^n\exp(4\pi \i k/N))(1-q^n\exp(-4\pi\i k/N)\right)\eta^6(\tau)}.\ee
So far everything is just the reciprocal of what we would get for fermions with the same twist angles.

However, for the zero-modes the story will be different.\footnote{This is inevitable, because the path integral for the bosonic fields $X^i$, in the absence of a worldsheet $B$-field, is
manifestly positive, while the integral over the fermion zero-modes gave a result that is not positive-definite.} 
For open strings ending on the Dp-brane, the fields $X_{p+2},\cdots, X_{10}$ have no zero-modes.   The fields $X_3,\cdots, X_{p+1}$ have untwisted zero-modes, as if we were considering
 open strings in $\R^{10}$.   And the fields $X_1,X_2$ have zero-modes in the $\sigma_1$ direction that are twisted by an angle $4\pi k/N$ in the $\sigma_2$ direction; we call these modes
 twisted zero-modes.
 
 The action for free bosons $X^i$ on a Euclidean string worldsheet, in conformal gauge, is
 \be\label{ofo} I = \frac{1}{4\pi\alpha'} \int \d\sigma_1\d\sigma_2 \sum_{i,\alpha}\left(\frac{\partial X^i}{\partial \sigma^\alpha}\right)^2. \ee
 The bosonic zero-modes are functions of the worldsheet ``time'' coordinate $\sigma_2$ only.   Integrating over $\sigma_1$, which runs from 0 to $\pi$, and writing $t$ for $\sigma_2$,
 we get an effective action for the zero-modes,
 \be\label{noffo}I=\frac{1}{4\alpha'}\int \d t \sum_i \left(\frac{\d X^i}{\d t}\right)^2. \ee
 The corresponding Hamiltonian is $H=\alpha' \sum_i P_i^2$, where $P_i=-\i \frac{\partial}{\partial X^i}. $
  The path integral with the action (\ref{noffo}) on a line segment of length $2\pi T$ (rather than a circle) with initial and final values $\vec X$ and $\vec X'$ at the two ends of the interval,
 computes the heat kernel
 \be\label{lofox}\la \vec X'|\exp(-2\pi T\alpha' \vec P^2)|\vec X\ra = \frac{1}{(8\pi^2\alpha' T)^{n/2}}\exp(-|\vec X - \vec X'|^2/8\pi\alpha' T), \ee
 where $n$ is the number of components of $\vec X$ and $\vec X'$ that are considered.
For open strings ending on the Dp-brane, there are
  $p-1$ zero-mode  components with no twist in the $t$ direction, so for those modes, $n=p-1$.  To get the path integral on the circle for these modes, we set $\vec X'=\vec X$ and integrate over $\vec X$.
The integral gives $V$, the volume of that intersection, times 
$ 1/(8\pi^2\alpha' T)^{(p-1)/2}$.    The result is then a factor
\be\label{toofo} \frac{V}{(8\pi^2\alpha' T)^{(p-1)/2}}.\ee
If we take the $\R^2/\Z_N\times \R^8$ model literally, then $V$ is infinite.   We could make $V$ finite  by toroidal compactification of the coordinates $X_2,\cdots, X_{p+1}$,
without significantly altering the rest of the  discussion. In practice, it is not necessary to do this explicitly.
At any rate, we are interested in identifying contributions (to the entropy, or to $\log \Tr\,\rho^N$) that
are proportional to $V$.   If we were studying a D9-brane, $V$ would be the volume of the Rindler horizon; for a more general Dp-brane, it is the volume of the brane worldvolume
intersected with the Rindler horizon.

Now let us consider the twisted zero-modes.   In going around the circle, $X_1$ and $X_2$ are twisted by the rotation matrix
 \be\label{nofox} R=\begin{pmatrix} \cos (4\pi k/N)& \sin (4\pi k/N)\cr -\sin (4\pi k/N)& \cos (4\pi k/N)\end{pmatrix}. \ee
The path integral on the circle is then given by setting $X'=R X$ and $n=2$ in eqn. (\ref{lofox}) and integrating over $X$, giving
 \be\label{zofo}\frac{1}{8\pi^2\alpha' T}\int \d^2\vec X \exp(-|(1-R) X'|^2/8\pi\alpha' T)=\frac{1}{|\det(1-R)|}.\ee
   Evaluating the determinant, we find that
 \be\label{trofo} \frac{1}{|\det(1-R)|}=\frac{1}{4\sin^2(2\pi k/N)}. \ee
This does not coincide with the inverse of the zero-mode contribution to a fermionic integral with the same twist angle.   From eqn. (\ref{lofo}), but with $k/N$ replaced by $2k/N$,
we see that the zero modes of a complex fermion with twist angle $4\pi k/N$ would contribute a factor $\pm 2\i \sin(2\pi k/N)$ in the path integral.   The zero modes of a twisted
boson contribute the absolute value squared of the inverse of this.    Thus there is an extra factor of $1/(\mp 2\i \sin(2\pi k/N))=\pm \i/2\sin(2\pi k/N)$ 
compared to what we would expect if the boson path integral
were just the inverse of a corresponding fermion path integral.

Putting all these factors together, the bosonic partition function for $k\not=0$ is
\be\label{zlofo} Z^B_{k,N}= \frac{V}{(8\pi^2\alpha' T)^{(p-1)/2}} \frac{1}{4\sin^2(2\pi k/N) q^{1/12}\prod_{n=1}^\infty(1-q^n\exp(4\pi\i k/N)(1-q^n \exp(-4\pi\i k/N))}\frac{1}{\eta^6(\tau)}.\ee
This can be written more simply in terms of the function $F(z,\tau)$ introduced in eqn. (\ref{nuco}):
\be\label{plofo}Z^B_{k,N}= \frac{V}{(8\pi^2\alpha' T)^{(p-1)/2}}\frac{\i}{2\sin 2\pi z}\frac{1}{F(2z,\tau)\eta^6(\tau)}.\ee

The overall partition function $Z_{k,N}=Z^F_{k,N}Z^B_{k,N}$ is then
\be\label{rofo}Z_{k,N}(\tau)=\frac{V}{(8\pi^2\alpha' T)^{(p-1)/2}}\frac{\i}{2\sin 2\pi z}\frac{F^4(z,\tau)}{F(2z,\tau)\eta^6(\tau)},~~~~z=k/N. \ee
The right hand side is the function $J(z,\tau)$ that we need in eqn. (\ref{zunno}) for the analytically continued partition function.

This function has the properties that were claimed in section \ref{strategy}.   To understand it better, let us look more closely at the function $F(z,\tau)$ defined in eqn. (\ref{nuco}).
The infinite product is absolutely convergent for $|q|<1$, so 
 $F(z,\tau)$ is holomorphic throughout that region.
There is an obvious antiperiodicity $F(z+1,\tau)=-F(z,\tau)$  (which leads to periodicity of $F^4(z,\tau)/F(2z,\tau)$), 
but actually there is a second periodicity under $z\to z+\tau$, which corresponds to $w\to qw$, with $w=\exp(2\pi\i z)$.
To prove this, we can freely rearrange terms in the product that defines $F(z,\tau)$, since this product is absolutely convergent, and we find
\be\label{tuclo}F(z+\tau,\tau)=F(z,\tau)\frac{(1-w^{-1})(q^{1/2}w^{1/2}-q^{-1/2}w^{-1/2})}{(1-qw)(w^{1/2}-w^{-1/2})}=F(z,\tau) q^{-1/2}w^{-1}.\ee
It follows then that
\be\label{pluco} F(2z+2\tau,\tau)=F(2z+\tau,\tau)q^{-1/2}(qw^2)^{-1}=F(2z,\tau) q^{-2}w^{-4}.\ee
So $F^4(z,\tau)/F(2z,\tau)$ is invariant under $z\to z+\tau$.    In other words,  as a function of $z$,  this function is doubly-periodic under the lattice shifts
$z\to z+m+n\tau$.

It is convenient to define
\be\label{nluco} G(z,\tau)=\frac{F^4(z,\tau)}{F(2z,\tau) \eta^6(\tau)}. \ee
The function $J(z,\tau)$ is
\be\label{tuco}J(z,\tau) = \frac{V}{(8\pi^2\alpha' T)^{(p-1)/2}}  \frac{\i}{2\sin 2\pi z} G(z,\tau).\ee 
The function $G(z,\tau)$  is doubly-periodic, meaning that for fixed $\tau$, $G(z,\tau)$ is a meromorphic function on the elliptic curve (or
genus 1 Riemann surface)   $E=\CC/(\Z+ \Z \tau)$.    From the product representation (\ref{nuco}), we see
that $F(z,\tau)$ vanishes if and only if $q^nw=1$ for some integer $n$, which means that $z$ is an element of the lattice $\Z+ \Z\tau$.  Similarly, $F(2z,\tau)$ vanishes if and only if 
$q^nw^2=1$ for some $n$, which means that $2z$ is an element of the lattice.    For $2z$ to be a lattice element means that $z$ is congruent, modulo a lattice shift, to one
of the half-lattice points $0,1/2,\tau/2$ or $(1+\tau)/2$.
Combining these statements, 
$G(z,\tau)$ has simple poles at the non-zero half-lattice points
$z=1/2, \tau/2,$ and $(1+\tau)/2$, 
coming from zeros of $F(2z,\tau)$, and it has a third order zero at $z=0$.   Up to a constant multiple, $G(z,\tau)$ is uniquely determined, for fixed $q$, by these poles and zeros,
and in particular the fact that the zero at $z=0$ is of odd order implies that $G(z,\tau)$ is an odd function of $z$:
\be\label{bruco} G(-z,\tau)=-G(z,\tau). \ee

Analogous properties of the partition function $Z(z,\tau)$ follow immediately, taking into account the additional factor of $1/\sin 2\pi z$. 
In particular, $J(z,\tau)$ has precisely the same poles as $G(z,\tau)$ except that its poles at $z=1/2$ mod $\Z$ are double poles.
Moreover
  \be\label{zuccot}J(z,\tau)=J(-z,\tau)=J(z+1,\tau). \ee

It is possible to write $G(q,z)$ in terms of theta functions, together with the Dedekind eta function \cite{Dabholkar, Takayanagi}. This representation is related to the product
formula that we have given by the Jacobi triple product formula.  Because $G(q,z)$ is a doubly-periodic
meromorphic function, it is also possible to write it as a lattice sum; see Appendix \ref{pfunction}.

\subsection{Residues}\label{Second}

In view of eqn. (\ref{tuco}), we can slightly rewrite the formula (\ref{zunno}) for the analytically continued partition function $Z_N(\tau)$ in terms of residues of $G(z,\tau)$ rather
than $J(z,\tau)$:
\begin{align}\label{goodone} \frac{2(8\pi^2\alpha'T)^{(p-1)/2}}{V} Z_N(\tau)=&-\frac{1}{N}\sum_{z_0\in\S_0} \frac{\i K_1(z_0,N)}{\sin 2\pi z_0} \Res_{z_0}G(z,\tau) 
-\frac{1}{N}\sum_{z_0\in\S'_1}\frac{\i K_2(z_0,N)}{\sin 2\pi z_0}\Res_{z_0} G(z,\tau)\cr &-\frac{1}{2\pi N}\left.\partial_z K_2(z,N)\right|_{z=1/2} \,\Res_{z=1/2} G(z,\tau).\end{align}
To evaluate this, we need a formula for  the residues of  $G(z,\tau)$.

We first consider the residue at $z=\tau/2$ modulo the lattice.     This corresponds to $w=q^{1/2}$, where $w=\exp(2\pi\i z)$.  From the product formula for $G(z,\tau)$, one finds that 
\be\label{otto} G(z,\tau)\widesim{z\to\tau/2} -\frac{1}{1-qw^{-2}}\cdot \frac{q^{-1/6}\prod_{n=1}^\infty (1-q^{n-1/2})^8}{\eta^8(\tau)}. \ee
Hence the residue of $G(z,\tau)$ at $z=\tau/2$ is\footnote{The residue is taken in the $z$ variable, 
that is we want the residue of the one-form $\d z \,G(z,\tau)$.} 
\be\label{zotto}\frac{\i}{4\pi} \frac{q^{-1/6}\prod_{n=1}^\infty (1-q^{n-1/2})^8}{\eta^8(\tau)}=\frac{\i}{4\pi}\frac{\eta^8(\tau/2)}{\eta^{16}(\tau)}. \ee

The other cases can be treated similarly.   
We note that $z=(1+\tau)/2$ modulo the lattice corresponds to $w=-q^{1/2}$.   We have
\be\label{lotto} G(z,\tau)\widesim{z\to(1+\tau)/2} \frac{1}{1-qw^{-2}}\cdot \frac{q^{-1/6}\prod_{n=1}^\infty (1+q^{n-1/2})^8}{\eta(\tau)^8}, \ee
so the residue of $G$ at $z=(1+\tau)/2$ is
\be\label{mottom}-\frac{\i}{4\pi} \frac{q^{-1/6}\prod_{n=1}^\infty (1+q^{n-1/2})^8}{\eta^8(\tau)}=-\frac{\i}{4\pi}\frac{\eta^8(\tau)}{\eta^8(2\tau)\eta^8(\tau/2)}. \ee

 Finally, let us look at the residue at $z=1/2$ modulo the lattice, which corresponds to $w=-1$.   We have   
 \be\label{jotto} G(z,\tau)\widesim{z\to 1/2} -\frac{1}{w-w^{-1}}\cdot \frac{16 q^{1/3}\prod_{n=1}^\infty (1+q^n)^8}{\eta^8(\tau)}, \ee
 so the residue of $G$ at $z=1/2$ is
 \be\label{ijotto}\frac{\i}{4\pi} \frac{16 q^{1/3}\prod_{n=1}^\infty (1+q^n)^8}{\eta^8(\tau)}= \frac{\i}{4\pi} \frac{16\eta^8(2\tau)}{\eta^{16}(\tau)}.\ee

As a check on these computations, the vanishing of the sum of the residues of the doubly-periodic differential $G(z,\tau)\d z$ at its poles at $z=1/2, \tau/2$, and $(1+\tau)/2$
gives the identity
\begin{align}\label{gotto}\frac{q^{-1/6}\prod_{n=1}^\infty(1-q^{n-1/2})^8}{\eta^8(\tau)}- \frac{q^{-1/6}\prod_{n=1}^\infty(1+q^{n-1/2})^8}{\eta^8(\tau)}+\frac{16q^{1/3}\prod_{n=1}^\infty (1+q^n)^8}{\eta^8(\tau)}=0.\end{align}
This is the identity, due originally to Jacobi,
 that was used by Gliozzi, Olive, and Scherk \cite{GSO} to demonstrate supersymmetry of the open-string spectrum using RNS fermions in light-cone gauge.
The three terms are the contributions to the partition function of the string from the three even spin structures on a torus.   If we label a spin structure in the usual way
as $\pm \pm$, where the two signs correspond to periodic or antiperiodic boundary conditions in the $\sigma_1$ and $\sigma_2$ directions, respectively, then 
the three terms in eqn. (\ref{gotto}) correspond to the $-+$, $--$, and $+-$ spin structures.

Of course, in our calculation, we started with Green-Schwarz fermions, not RNS fermions.  But we introduced a $\Z_N$ twist to construct the orbifold, and the evaluation of the
partition functions in terms of residues involved analytically continuing with respect to the twist parameter and extracting residues that arise at the points at which the bosons are untwisted -- leading to a pole in $G(z,\tau)$ from a bosonic
zero-mode -- while the fermions are twisted by signs.   The twisting of the fermions determines an even spin structure and the residue of $G(z,\tau)$ at such a point is the RNS partition
function for that spin structure.

\subsection{Closed-String Sector}\label{closed}

For closed strings, the $\R^2/\Z_N\times \R^8$ orbifold has NS-NS tachyons  and massless Ramond-Ramond states in twisted sectors.  This was shown in \cite{Dabholkar}
using both the Green-Schwarz and RNS descriptions.

A short explanation in the Green-Schwarz language is as follows.  We parametrize a closed string by an angular variable $\sigma_1$, with $\sigma_1\cong \sigma_1+2\pi$.  
Consider closed strings that are twisted by an element $U^k\in\Z_N$ in the sense that, for any worldsheet
field $\Phi$, $\Phi(\sigma_1+2\pi)= U^k\Phi(\sigma_1)$.   Since the value of $k$ only matters mod $N$, we can assume that $-N/2<k<N/2$.   For $k>0$,
the partition function that counts the right-moving oscillator modes in this twisted sector is $-G((k/N)\tau,\tau)$. 
(For $k<0$,  the formula is $+G((k/N)\tau,\tau)$, leading to a similar analysis.)  To construct the orbifold
spectrum in the twisted sector, we have to combine the right-moving oscillator spectrum with a similar left-moving spectrum, and project onto the subspace of $\Z_N$ invariants.  
We have
\be\label{tuvu} -G((k/N)\tau,\tau)=q^{-k/N}-4+\mathcal{O}(q^{k/N}). \ee
The ground state, in RNS language, is an NS sector tachyon with $L_0=-k/N$, and the first excited level consists of 4 Ramond sector states with $L_0=0$.
 When we combine left- and right-movers and project onto the subspace of $\Z_N$ invariants, we are left with an NS-NS tachyon with $L_0=\bar L_0=-k/N$
 and $4^2=16$ Ramond-Ramond states with $L_0=\bar L_0=0$.   (The states with $L_0=-k/N$, $\bar L_0=0$, or vice-versa, are removed when one projects onto
 $\Z_N$ invariants.)   
 
In the annulus partition function, we expect to see closed-string states propagating in the crossed channel. 
Let us discuss the contribution that we should expect from the twisted RR states with
$L_0=\bar L_0=0$.  We consider a twisted sector closed-string state that is emitted on the left in fig. \ref{Annulus} of section \ref{partfn}, propagates
for a proper time $2\pi\t T$, and is absorbed on the right.    We have to compute the appropriate matrix element of $\exp(-2\pi H\t T)$, where $H=L_0+\bar L_0$ is
the Hamiltonian for the mode in question.     Since we consider a mode whose oscillator content has $L_0=\bar L_0=0$, $H$ receives a contribution only from the effective
Hamiltonian of the untwisted bosonic zero-modes.   This effective Hamiltonian can be worked out as in the discussion of eqn. (\ref{noffo}), with the only change
being $\alpha'\to \alpha'/2$ because we take the closed-string circumference to be $2\pi$.    Hence $H=\frac{\alpha'}{2}\sum_iP_i^2$, and the analog of
(\ref{lofox}) is 
\be\label{plofox} \la \vec X'|\exp(-\pi \t T\alpha' \vec P^2)|\vec X\ra = \frac{1}{(4\pi^2\alpha' \t T)^{4}}\exp(-|\vec X - \vec X'|^2/4\pi\alpha' \t T),\ee
which gives the amplitude for a twisted sector closed-string state to be emitted at a point $X$
 on the intersection of the Dp-brane worldvolume with the Rindler horizon and absorbed at a point $X'$ on that intersection.  
 (For closed strings, we have set $n=8$, since all 8 bosonic coordinates along the Rindler horizon have zero modes in the closed-string sector.)
  To get the full amplitude, we have to integrate  $X$ and $X'$ over the intersection of the Dp-brane worldvolume with the Rindler horizon. The dimension of this intersection is $p-1$,
  so that is the number of components of $X$ and $X'$ over which we integrate.  The integral over $X$ and $X'$ gives $V/(4\pi^2\alpha' \t T)^{(9-p)/2}$.  We also have to integrate
  over the proper time $\t T$.  The contribution of the massless modes to this integral behaves as 
 \be\label{kiofox} \frac{V}{(4\pi^2\alpha')^{(9-p)/2}}\int^\infty_{\cdots}\frac{\d \t T}{\t T^{(9-p)/2}},\ee 
 and this will be the dominant behavior for large $\t T$ if there are no tachyons.
We will look for such behavior in section \ref{comprho}.
 
 A twisted sector state in which the oscillator modes have $L_0=\t L_0=n$ would, by the same reasoning, give a contribution that would behave for large $\t T$
 as \be\label{wofox} \frac{V}{(4\pi^2\alpha')^{(9-p)/2}}\int^\infty_{\cdots}\frac{\d \t T}{\t T^{(9-p)/2}} \exp(-4\pi n\t T). \ee
 We will find such contributions in section \ref{comprho}, but after analytic continuation, we will also find terms that cannot quite be put in this form.

\section{Computation of $\Tr\,\rho^\N$}\label{comprho}

We now have the information we need to make explicit the formula (\ref{goodone}) for the analytically continued partition function $Z_N(\tau)$.
We work out first the contributions proportional to $K_1$ and then those proportional to $K_2$.

\subsection{The $K_1$ Contribution}\label{ramonds}

The contribution to the right hand side of eqn. (\ref{goodone}) that is proportional to $K_1$ is a sum over points in the cylinder $z\cong z+1$ that are congruent mod $\tau$
to $\tau/2$.   Those points are at $z=(r+1/2)\tau=(r+1/2)\i T$, $r\in \Z$.    A short calculation, using the residue formula (\ref{zotto}),
 shows that the contribution of these points to the right hand side of eqn. (\ref{goodone})
is
\be\label{poffo}-\frac{1}{4}\sum_{r\in\Z}\left(\coth \pi N(r+1/2)T -\frac{1}{N}\coth \pi(r+1/2)T\right) \frac{1}{\sinh 2\pi(r+1/2)T} \frac{\eta^8(\i T/2)}{\eta^{16}(\i T)}. \ee

Going back to eqns. (\ref{zoffo}) and (\ref{woffol}), the corresponding contribution to $\left.\log \Tr \,\rho^{1/N}\right|_1$ is
\be\label{zokan}-\frac{V}{8(8\pi^2\alpha')^{(p-1)/2}}\int_0^\infty\frac{\d T}{T^{(p+1)/2} }\sum_{r\in\Z}\left(\coth \pi N(r+1/2)T-\frac{1}{N}\coth \pi(r+1/2)T\right) \frac{1}{\sinh 2\pi(r+1/2)T} \frac{\eta^8(\i T/2)}{\eta^{16}(\i T)}. \ee

$T\to\infty$ is the infrared region in the open-string channel, the region in which one would expect to agree with a 1-loop field theory computation using the massless states of the open string.
The $T\to \infty$ behavior is dominated by $r+1/2=\pm 1/2$.   After taking account of  the large $T$ behavior of the eta function, $\eta(\i T)\sim \exp(-2\pi T/24)$,
one finds that the exponential factors cancel and
the  integral behaves for large $T$ as $\int^\infty_{\cdots}\d T/T^{(p+1)/2}$, and converges for $T\to\infty$ if $p>1$.    This is the expected behavior in field theory.
    For $p=1$, the integral is logarithmically divergent for $T\to\infty$.   This infrared divergence is present in the field theory limit (even for integer $N$)
and we cannot expect to get rid of it by going to string theory.

$T\to 0$ is the ultraviolet region from the point of view of field theory, but in string theory, as usual, one expects to describe the $T\to 0$ behavior in terms of tree-level closed-string exchange in the crossed
channel.  The proper time parameter in the closed-string channel is $\t T=1/2T$, as we recalled in the introduction to section \ref{partfn}.   To determine the behavior for small $T$,
as usual one uses the modular identity 
\be\label{okan}\eta(\i T)=\frac{1}{\sqrt T}\eta(\i/T), \ee
which implies
\be\label{wokan}\frac{\eta^8(\i T/2)}{\eta^{16}(\i T)}=(2T)^4 \frac{\eta^8(2\i/T)}{\eta^{16}(\i/T)}=(2T)^4\frac{\eta^8(4\i \t T)}{\eta^{16}(2\i \t T)}.\ee
So we can rewrite the integral (\ref{zokan}) in terms of $\t T$:
\begin{align}\label{qokan}-\frac{V}{8\cdot 2^{(9-p)/2}(8\pi^2\alpha')^{(p-1)/2}}\int_0^\infty\frac{\d \t T}{\t T^{(11-p)/2} }\sum_{r\in\Z}&\left(\coth (\pi N(r+1/2)/2\t T)-
\frac{1}{N}\coth (\pi(r+1/2)/2\t T)\right) \cr &\times \frac{1}{\sinh \pi(r+1/2)/\t T} \frac{16\eta^8(4\i \t T)}{\eta^{16}(2\i \t T)}. \end{align}

The sum over $r$ is of the general form  $\sum_{r\in\Z} W((r+1)/\t T)$ with a certain function $W$ that is holomorphic and bounded in a strip containing the real axis.   For such a function, it is useful
to write
\be\label{toombo} \sum_{r\in \Z}W((r+1/2)/\t T)=\sum_{m\in\Z} \int_{-\infty}^\infty \d r\,\exp(2\pi \i mr) W((r+1/2)/\t T)=\t T\sum_{m\in\Z}\int_{-\infty}^\infty \d x\,\exp(2\pi i m(x\t T-1/2)) W(x), \ee
where in the last step we change 
 variables from $r$ to $x=(r+1/2)/\t T$.  This procedure is essentially Poisson resummation.
The terms with $m\not=0$ vanish exponentially for $\t T\to\infty$, as one learns by deforming the integration contour above or below the real $r$ axis.  So modulo exponentially
small terms, we reduce to $\t T \int_{-\infty}^\infty \d x \,W(x)$.      In the present case,  this is $\t T Y(N)$ with
\be\label{lombo}Y(N)= \int_{-\infty}^\infty \d x\,\left(\left(\coth (\pi N x/2)-
\frac{1}{N}\coth (\pi x/2)\right) \frac{1}{\sinh \pi x} \right).\ee
There does not seem to be a convenient closed form for this integral, but some special cases relevant to the entropy are
\be\label{bombo}Y(1)=0,~~~Y'(1)=\frac {\pi}{4}-\frac{1}{\pi}. \ee

The large $\t T$ behavior of the integral (\ref{qokan}) is
\be\label{lokan}-\frac{V}{8\cdot 2^{(9-p)/2}(8\pi^2\alpha')^{(p-1)/2}}\int_{\cdots}^\infty\frac{\d \t T}{\t T^{(9-p)/2} } 16 Y(N). \ee
As discussed in section \ref{closed}, this is the expected behavior for tree level exchange of a massless twisted sector closed-string state.
The integral converges for $p\leq 6$.  Since $9-p$ is the codimension of the Dp-brane worldvolume in spacetime, the divergence for $p>6$ occurs
when this codimension is less than 3.   This divergence is similar to the standard infrared divergence that occurs for a Dp-brane of codimension less than 3, due
to tree-level exchange of massless dilatons and gravitons.    A Dp-brane of codimension less than 3 cannot really be treated as an isolated object, in a way that is possible
when the codimension is at least 3. 

Now let us discuss the exponentially small corrections for large $\t T$.   First of all, tree level exchange of a closed-string state is expected to produce a contribution proportional to 
$\exp(-2\pi \t T(L_0+\bar L_0))$ (times the amplitude for the state in question to be produced by the boundary on the left in fig. \ref{Annulus} of section \ref{partfn} and annihilated on the right).
We note that
\be\label{noat}\frac{16\eta^8(4\i \t T)}{\eta^{16}(2\i \t T)}=\frac{16 \prod_{n=1}^\infty (1+\exp(-4\pi n\t T))^8}{1-\exp(-4\pi n\t T)^8} \ee
is the same as $\Tr \exp(-4\pi  L_0)$ in the Ramond sector of open strings, or equivalently in the right-moving Ramond sector of closed strings.   This has a simple interpretation:
a standard fact in D-brane physics is that the closed-string states that propagate in the crossed channel of the annulus have the same Fock space structure for left- and right-movers,
so the enumeration of these states is the same as the enumeration of states in the right-moving sector, with the usual $\exp(-2\pi(L_0+\bar L_0)\t T)$ replaced  by $\exp(-4\pi L_0\t T)$, since  these
states have $L_0=\bar L_0$.    This suggests that the $K_1$ contribution to the analytically continued partition
function represents, for large $\t T$, the contribution of Ramond-Ramond states in the crossed channel.  (This interpretation may be oversimplified, as it does not take into account the
sum over $r$ in eqn. (\ref{qokan}), which also depends on on $\t T$.    If the interpretation is valid, it depends on the precise choice we made of the $K$ function with no pole at $z=0$; see the conclusion
of section \ref{strategy}.)  The massless twisted sector modes responsible
for the behavior in eqn. (\ref{lokan}) are presumably  simply the massless twisted sector Ramond-Ramond states of the orbifold; see section \ref{closed}.

Some exponentially small terms come from the expansion of eqn. (\ref{noat}) in powers of $\exp(-2\pi \t T)$.
Additional exponentially small contributions come from terms with $m\not=0$ in eqn. (\ref{toombo}), which we have omitted so far.  
The integral for general $m$ is
\be\label{innow} \int_{-\infty}^\infty \d x\,\exp(2\pi \i m(x\t T-1/2))\left(\left(\coth (\pi N x/2)-
\frac{1}{N}\coth (\pi x/2)\right) \frac{1}{\sinh \pi x} \right).\ee
Let us first suppose that $N$ is an odd integer.    The singularities of the integrand are then simple poles at $x=2\i k/N$, for nonzero integer $k$ not divisible by $N$.   The contribution of a simple pole
at $\Im\,x = b$ is evaluated by shifting the integration contour upward or downward in the complex $x$-plane depending on the sign of $mb$.    This contribution is a constant multiple of
$\exp(-2\pi |mb|\t T)$.     With $b=2k/N$, this is $\exp(-4\pi( k'/N) \t T)$, where $k'=|mk|$.   That is the expected behavior due to twisted sector states with $L_0=\bar L_0=k'/N$, which of course
are present in the orbifold theory at levels of positive mass.

What happens if we analytically continue away from odd integer $N$? The most important difference is that now there are double poles at $x=2\i k$ for integer $k$, in addition
to the simple poles.
The contribution of such a double pole is not a simple exponential. The residue at a double pole has an extra factor of $\t T$, leading to a contribution of the
 form $\t T \exp(-4\pi k'\t T)$, again with $k'=|mk|$.   What is the meaning of the extra factor of $\t T$?

It is not possible to get such an extra factor in a unitary quantum field theory, in which $L_0$ and $\t L_0$ can be diagonalized.   However, such contributions are possible in a logarithmic
conformal field theory \cite{RS,G,C}.  In such theories, which are necessarily nonunitary, $L_0$ and $\bar L_0$ are not diagonalizable.   Suppose that in a two-dimensional subspace,
\be\label{wonk}L_0 = k' I_2 +\begin{pmatrix} 0 & 1\cr 0 & 0 \end{pmatrix},\ee
where $I_2$ is the $2\times 2$ identity matrix.    Then in this subspace
\be\label{lonk} \exp(-4\pi \t TL_0) =\exp(-4\pi k' \t T)\begin{pmatrix} 1 & -4\pi \t T\cr 0 & 1\end{pmatrix}. \ee
If the boundary state associated to the Dp-brane, when projected to this two-dimensional subspace, is a generic linear combination of the two states, then the contribution of these
states in the closed-string channel of the annulus will have a term proportional to $\t T \exp(-4\pi k'\t T)$.

Thus there is an indication that analytic continuation away from integer $N$, if it can be carried out systematically,  gives a logarithmic conformal field theory.    For the $K_1$ contribution, which we tentatively interpret
as the Ramond-Ramond contribution in the closed-string channel, we have found such an indication  only for massive modes.   For the $K_2$ contribution,
we will find such behavior also for massless states.

\subsection{The $K_2$ Contribution}\label{nss}

The contribution to the right hand side of eqn. (\ref{goodone}) that is proportional to $K_2$ is a sum over points in the cylinder $z\cong z+1$ that are congruent mod $\tau$
to either $(1+\tau)/2$ or $1/2$.    Let us first consider the contributions at $z=1/2+(r+1/2)\tau$, with integer $r$.
Using the residue formula (\ref{mottom}),
one finds that the contribution of these points to the right hand side of eqn. (\ref{goodone})
is
\be\label{poffor}-\frac{1}{4}\sum_{r\in\Z}\left(\tanh \pi N(r+1/2)T-\frac{1}{N}\tanh \pi(r+1/2)T\right) \frac{1}{\sinh 2\pi(r+1/2)T} \frac{\eta^8(\i T)}{\eta^{8}(\i T/2)\eta^8(2\i T)}. \ee
The contributions at $z=1/2+r\tau$ are similar; we just replace $r+1/2$ with $r$ and use the residue formula (\ref{ijotto}), to get
\be\label{woffor}\frac{1}{4}\sum_{r\in\Z}\left(\tanh \pi NrT-\frac{1}{N}\tanh \pi rT\right) \frac{1}{\sinh 2\pi rT}\frac{16\eta^8(2\i T)}{\eta^{16}(\i T)}. \ee
(The $r=0$ term here actually comes from the last term in eqn. (\ref{goodone}), which originates in the double pole of the partition function at $z=1/2$.)

The infrared limit $T\to\infty$ in the open-string channel comes from the terms in eqn. (\ref{poffor}) with $r+1/2=\pm 1/2$ and the $r=0$ contribution in eqn. (\ref{woffor}).   Other contributions vanish exponentially.

Let us look at the opposite limit $T\to 0$.   We have with the help of eqn. (\ref{okan}) 
\be\label{lobel}\frac{\eta^8(\i T)}{\eta^{8}(\i T/2)\eta^8(2\i T)}=T^4\frac{\eta^8(2\i \t T)}{\eta^{8}(\i \t T)\eta^8(4\i \t T)},~~~~\frac{16\eta^8(2\i T)}{\eta^{16}(\i T)}=T^4\frac{\eta^8(\i \t T)}{\eta^{16}(2\i\t T)}.\ee
The analog of eqn. (\ref{qokan}) is 
\begin{align}\label{qolkan}-\frac{V}{8\cdot 2^{(9-p)/2}(8\pi^2\alpha')^{(p-1)/2}}\int_0^\infty\frac{\d \t T}{\t T^{(11-p)/2} }\left( A(\t T)\frac{\eta^8(2\i \t T)}{\eta^{8}(\i \t T)\eta^8(4\i \t T)}
+ B(\t T) \frac{\eta^8(\i \t T)}{\eta^{16}(2\i\t T)}\right) , \end{align}
with
\begin{align}\label{olkan} A(\t T)& = -\frac{1}{4}\sum_{r\in\Z}\left(\tanh \pi N(r+1/2)/2\t T -\frac{1}{N}\tanh \pi(r+1/2)/2 \t T\right) \frac{1}{\sinh \pi(r+1/2)/\t T}\cr
                                        B(\t T)& = \frac{1}{4}\sum_{r\in\Z}\left(\tanh \pi Nr/2 \t T-\frac{1}{N}\tanh \pi r/2 \t T\right) \frac{1}{\sinh \pi r/\t T}.\end{align}

The functions
\begin{align}\label{flossy} \frac{\eta^8(2\i \t T)}{\eta^{8}(\i \t T)\eta^8(4\i \t T)}& = \exp(2\pi\t T)\frac{\prod_{n=1}^\infty (1+\exp(-2\pi(2n-1)\t T))^8}{(1-\exp(4\pi n\t T)) ^8}=\exp(2\pi \t T) + 8+{\mathcal O}(\exp(-2\pi \t T))\cr
    \frac{\eta^8(\i \t T)}{\eta^{16}(2\i\t T)}&=   \exp(2\pi\t T)\frac{\prod_{n=1}^\infty (1-\exp(-2\pi(2n-1)\t T))^8}{(1-\exp(4\pi n\t T)) ^8}=\exp(2\pi \t T) - 8+{\mathcal O}(\exp(-2\pi \t T))\end{align}
    can be interpreted, respectively, as $\Tr \,\exp(-4\pi L_0 T)$ and as $\Tr\, (-1)^F \exp(-4\pi L_0 T)$, in the right-moving Neveu-Schwarz sector of a closed string.  This suggests
that the $K_2$ contribution to the partition function represents, for large $\t T$, the contribution of NS - NS states in the crossed channel.    (Like the analogous comment in section
\ref{ramonds}, this interpretation may be oversimplified.)

The contributions that can potentially grow exponentially for large $\t T$ come from the $\exp(2\pi\t T)$ terms on the right hand side of eqn. (\ref{flossy}).  The corresponding contributions to the $A$ and $B$
terms in  eqn. (\ref{qolkan}), after setting $r=s/2$, can be conveniently combined together and written 
\begin{equation}\label{goffo}\frac{1}{4} \exp(2\pi\t T) \sum_{s\in\Z}(-1)^s \left(\tanh \pi Ns /4\t T-\frac{1}{N}\tanh \pi s/4\t T\right) \frac{1}{\sinh \pi s/2\t T}.\ee

We will return to this sum in a moment, but first let us look at the contributions that come from the constant terms on the right hand side of eqn. (\ref{flossy}).   One might expect that they correspond
to the exchange of massless closed-string states.   Now the $A$ and $B$ terms combine to
\be\label{loffo}-2 \sum_{s\in \Z}\left(\tanh \pi Ns /4\t T-\frac{1}{N}\tanh \pi s/4\t T\right) \frac{1}{\sinh \pi s/2\t T}.\ee
The large $\t T$ behavior of this sum can be analyzed by a similar method to what we used for the sum that we encountered in eqn. (\ref{qokan}).
Modulo exponentially small terms, the sum can be replaced by an integral
\be\label{noffor}-2\t T \int_{-\infty}^\infty\d x \left(\tanh \pi Nx/4-\frac{1}{N}\tanh \pi x/4\right) \frac{1}{\sinh \pi x/2} , \ee
leading again to a contribution to the $\t T$ integral in eqn. (\ref{qolkan}) of the form $\int_{\cdots}^\infty \d \t T/\t T^{(9-p)/2}$.   
There are also exponentially small corrections, controlled by poles of the function
\be\label{milok} V(x)= \left(\tanh \pi Nx/4-\frac{1}{N}\tanh \pi x/4\right) \frac{1}{\sinh \pi x/2} .\ee
For odd integer $N$, this function has simple poles, corresponding  to contributions of closed-string states with $L_0>0$.
Upon continuation away from  odd integer $N$, $V(x)$ has double poles, suggesting a nondiagonalizability of the spectrum, again at $L_0>0$, as we found in section \ref{ramonds}.

Going back to the potentially tachyonic sum in eqn. (\ref{goffo}), Poisson resummation converts it to 
\begin{align}\label{toffo}\frac{1}{4} &\exp(2\pi\t T) \sum_{m\in\Z}\int_{-\infty}^\infty \d s \exp(\i \pi (2m+1)s) \left(\tanh \pi Ns /4\t T-\frac{1}{N}\tanh \pi s/4\t T\right) \frac{1}{\sinh \pi s/2\t T}
\cr = &\frac{\t T}{2} \exp(2\pi\t T) \sum_{m\in\Z}\int_{-\infty}^\infty \d x \exp(2\i \pi (2m+1)x\t T) \left(\tanh \pi Nx/2-\frac{1}{N}\tanh \pi x/2\right) \frac{1}{\sinh \pi x},
\end{align} where $x=s/2 \t T$.
All contributions to the integral are exponentially small for large $\t T$.   The dominant contributions come from $2m+1=\pm 1$ and from the poles closest to the real axis of the function
\be\label{lome}P(x)= \left(\tanh \pi Nx/2-\frac{1}{N}\tanh \pi x/2\right) \frac{1}{\sinh \pi x}.\ee

Let us first discuss the situation for odd integer $N$.   All poles of $P(x)$ are simple poles. 
The poles closest to the real axis are at $x=\pm \i /N$.   But we can get a somewhat clearer picture by considering all of the poles for $ |\Im \,x|\leq 1$. As we will see in a moment, poles
at $|\Im\, x|<1$ 
  make exponentially growing contributions to the analytically continued path integral, while a pole at  $|\Im\, x|=1$ will make a contribution with a constant or polynomial
 dependence on $\t T$.     For odd integer $N$, the poles with $|\Im\, x|\leq 1$ are  simple poles 
at $x=\pm \i\left(1-2k/N\right)$, $k=1,\cdots, (N-1)/2N$.    After shifting the contour in the direction of positive or negative $\Im\,x$, we find that the contribution of such
a pole to the integral in eqn. (\ref{toffo}) is a constant times $\exp(-2\pi |x|\t T)=\exp(-2\pi\t T(1-2k/N))$.    Including the prefactor $\exp(2\pi\t T)$ in eqn. (\ref{toffo}), the contribution of such a pole
to the analytically continued partition function is proportional to $\exp(4\pi (k/N)\t T)$.     This is $\exp(-4\pi \t T L_0)$, with $L_0=-k/N$.   That is the appropriate value for the
tachyonic twisted sector closed-string modes, as reviewed in section \ref{closed}.

What happens when we analytically continue away from odd integer $N$?   There is then a double pole at $x=\pm \i$, so the closed-string spectrum apparently becomes undiagonalizable.
The double pole makes a contribution $\t T\exp(-2\pi\t T)$ to the integral.  Combining this with the explicit factor $\t T \exp(2\pi \t T)$ that is already present in eqn. (\ref{toffo}), we find that
the double pole contributes a term $\t T^2$ to the expression that appears in parentheses in the formula (\ref{qolkan}) for the $K_2$ contribution to the analytically continued partition function.
This leads to an integral over $\t T$ whose large $\t T$ behavior is $\int^\infty_{\cdots}\d \t T/\t T^{(7-p)/2}$, not the familiar $\int^\infty_{\cdots}\d \t T/\t T^{(9-p)/2}$.

As long as $\Re\,1/N<1$, this term is subleading relative to exponentially growing contributions from poles with $|\Im \,x|<1$.    However, as soon as $\Re\,{1/N}>1$, there are no poles for
$|\Im\,x|<1$ and the poles with the smallest imaginary part are the double poles at $x=\pm \i$.  (This reflects the fact that for $\Re\,1/N>1$, the function $K_2(z,N)$ has no pole
with $0< \Re\,z<1$.   See eqn. (\ref{holfe}).)    Thus, as stated in the introduction, when continued to $\Re\,1/N>1$ or equivalently
$\Re\,\N>1$ where $\N=1/N$, the theory appears to be nontachyonic.    In this range of $\N$, the dominant contribution for large $\t T$ comes from the double pole at $x=\pm \i$
and has an extra and difficult to interpret factor of $\t T$ relative to what one would get from conventional massless closed-string exchange.

As in eqn. (\ref{wonk}), the double pole may correspond to a two-dimensional space of closed-string states that appears when $N$ is not an odd integer, and in which
\be\label{unb}L_0=\begin{pmatrix} 0 & 1\cr 0 & 0 \end{pmatrix}. \ee
These are states that propagate only on the Rindler horizon (analogous to twisted sector states of the orbifold) since their contribution is proportional to the volume $V$ of the intersection
of the Dp-brane worldvolume with the Rindler horizon, and not to the full Dp-brane worldvolume.     If it is true that the $K_2$ contribution corresponds in the closed-string channel to the NS-NS
sector, then the two states in question are NS-NS states.

To compute the 1-loop contribution to the entropy, we have to differentiate the 1-loop partition function 
with respect to $N$ at $N=1$.  Differentiating the integral in (\ref{toffo}) with respect to $N$ and setting $N=1$, we get
\be\label{voffo}
\int_{-\infty}^\infty \d x \exp(2\i \pi (2m+1)x\t T) \left(\frac{\pi x/2}{\cosh^2 \pi x/2}+ \tanh \pi x/2\right) \frac{1}{\sinh \pi x} .\ee
Now the integrand has triple poles at $x=\pm \i$.   The residue of a triple pole contributes an extra factor of $\t T^2$, so that the integral in (\ref{voffo}) behaves as $\t T^2\exp(-2\pi\t T)$ for large $\t T$.
This means that the integral controlling the entropy behaves for large $\t T$ as $\int^\infty_{\cdots}\d \t T/\t T^{(5-p)/2}$.   
For the $K_1$ contribution discussed in section \ref{ramonds}, differentiating with respect to $N$ at $N=1$ gives the expected behavior, with a coefficient that is determined by eqn. (\ref{bombo}).

Concretely, the integrand in eqn. (\ref{toffo}) has a double pole at $x=\i$ and a simple pole at $x=\i/N$.   For $N$ near 1 and $x$ near $\i$, the integrand in (\ref{toffo}) behaves as 
\be\label{wonder} \frac{1}{(x-\i)^2}-\frac{1}{(x-\i)(x-\i/N)}= \frac{\i(1-1/N)}{(x-\i)^2(x-\i/N)}. \ee
For generic $N$ near 1, there is a double pole at $x=\i$ and a simple pole at $x=\i/N$.   At $N=1$, both of these poles disappear, but if one differentiates with respect to $N$ before
setting $N=1$, one is left with a triple pole at $x=\i$.     To describe this situation requires a three-dimensional space
of states, in which $L_0$ can look like
\be\label{looklike}\begin{pmatrix}1-1/N & 0 & 1\cr 0 & 0 & 1\cr 0 & 0 & 0 \end{pmatrix}.\ee
This will reproduce the large $\t T$ behavior we have found, in all regimes.

The extra factors of $\t T$ and $\t T^2$ that we have obtained from double and triple poles at $x=\pm \i$ are difficult to interpret.   Taking what we have found at face value, it appears
that the 1-loop contribution to $\log \Tr\,\rho^\N$, $\Re\,\N>1$ from a Dp-brane crossing the horizon is finite only for $p\leq 4$, while the corresponding 1-loop contribution to the entropy
is finite only for $p\leq 2$.  One would have expected the bound $p\leq 6$ from tree-level exchange of a massless closed string.    The proper interpretation is far from clear.

One point worth mentioning is the following.   The Rindler horizon is flat in the absence of a Dp-brane, but the gravitational
field of a Dp-brane crossing the Rindler horizon modifies this flat metric.   The correction to the volume form of the Rindler horizon is proportional to $1/|x|^{7-p}$ where $|x|$ is the distance from the Dp-brane.
The integral of this over $9-p$ transverse directions behaves as $\int^\infty_{\cdots}  \d^{9-p}x \,|x|^{-(7-p)}$ and is infrared divergent.   Thus the Dp-brane corrects the classical Bekenstein-Hawking contribution
to the Rindler entropy -- the volume of the horizon -- by an infrared-divergent amount.  In string theory, this might appear as a disc contribution, though the proper machinery for that
computation is not currently known.   If the disc contribution to the open-string entropy is infrared divergent, perhaps one should not be too surprised to have such a divergence in the annulus contribution.
Still, the rationale for the precise power-law behavior that occurs at large $\t T$ is not clear.

\vskip 1cm \noindent
{\it Acknowledgements}  I thank  V. Rosenhaus, P. Sarnak, and especially A. Dabholkar for discussions concerning this topic.    Research partially supported by NSF Grant PHY-1606531. 
\vskip 1cm

\begin{appendix}
\section{A Reformulation Of The Problem Of Analytic Continuation}\label{reform}

Let us first recall Buscher's path integral derivation of $T$-duality \cite{Buscher}.   We start with some two-dimensional
theory with a U(1) global
symmetry and an action that we will schematically call $I(X)$, with bosonic fields  $X$, some of which are charged under U(1)
(there may also be fermion fields).   The first step is to gauge the symmetry,
adding a U(1) gauge field $A$, with field strength $F=\d A$,
 and to promote $I(X)$ to a gauge-invariant action $I(X,A)$.    Then we add a circle-valued field
$B$, with period $2\pi$, and consider the extended action on a two-manifold $\Sigma$:
\be\label{toggo}\hat I(X,A,B)=\i\int_\Sigma \frac{BF}{2\pi}+I(X,A). \ee
The extended theory with this action is precisely equivalent to the original theory.   To see this, one first performs
the path integral over $B$.   If $B$ were a real-valued field, the path integral over $B$ would give a delta function setting
$F=0$, and nothing more, meaning that the connection $A$ would be flat, but not necessarily trivial.   Because $B$ is circle-valued, one also has
to sum over winding modes of $B$.   This gives a further delta function forcing the global holonomies of $A$ to be trivial.  This
statement depends on the precise normalization of the $BF$ term in the action, and will change when this coefficient is changed
(as we do below).
With the chosen coefficient, after summing over winding modes, we learn that
 $A$ is gauge-equivalent to 0.  So one can go to the gauge $A=0$, and one then gets back the original theory with action $I(X)$.

On the other hand, given that the fields $X$ include scalar fields charged under U(1),
one can go to a
 ``unitary gauge''  (at least in a suitable portion of field space) in which the phase of some charged field\footnote{For a 
 complete gauge fixing of this type to be possible, the action of U(1) on the target space of the original theory should
be faithful, meaning in the case of a linear model that the U(1) charges of the original fields should be relatively prime.  Then some product of those fields
has charge 1, and the unitary gauge is defined by requiring this product to be positive.}
is set to 0, leaving a reduced set of fields $X'$.  In such a unitary gauge (assuming the original action $I(X)$ is quadratic
in first derivatives)
$A$  has a nondegenerate kinetic energy without
derivatives and can be integrated out in a local fashion.   This gives a new action $\t I(X',B)$, which describes the $T$-dual
of the original theory.

A fairly well-known variant of this arises if we modify the coefficient of the $BF$ term in the action and consider instead
\be\label{poggo} \hat I_N(X,A,B)=\i N\int_\Sigma \frac{BF}{2\pi}+I(X,A). \ee
Here, since $B$ has period $2\pi$, and the flux of $F$ is an integer multiple of $2\pi$, the coefficient $N$ must be an integer in order that the action is well-defined mod $2\pi\i$,
as is necessary in order for the Feynman path integral of the theory to make sense.  Repeating the derivation, the integral over 
small fluctuations of $B$ still gives a delta function setting $F=0$, but the sum over winding modes of $B$ now gives a delta
function telling us that the holonomies of $NA$  are trivial, not necessarily the holonomies of $A$.   This means that $A$ reduces to a $\Z_N$
gauge field and the theory is equivalent to a $\Z_N$ gauge theory, with only a $\Z_N$ subgroup of the original U(1) global symmetry being gauged.
  This $\Z_N$ gauge theory  is
alternatively called a $\Z_N$ orbifold of the original theory.   The path integral of the $\Z_N$ gauge theory consists of summing over
$\Z_N$ twists around all cycles in $\Sigma$ and dividing by an overall factor of $N$ (this last step is the $\Z_N$ version of Faddeev-Popov gauge fixing).
That is the usual recipe for a $\Z_N$ orbifold \cite{orbi}.

So a restatement of the problem of analytically continuing the orbifold with respect to $N$ is to analytically continue the theory with action $\hat I_N(X,A,B)$
with respect to $N$.  Unfortunately, it is not clear how to proceed.  The problem is somewhat similar to the problem of analytically continuing three-dimensional
Chern-Simons gauge theory with respect to its integer coupling constant $k$.  That problem actually has an answer that is useful for some purposes \cite{WittenCS},
but what is known in Chern-Simons theory does not have an obviously useful analog for our present problem.  

Going back to eqn. (\ref{toggo}), one  can ask  what would be  an orbifold in the dual description.\footnote{This paragraph is a response to a comment
by J. Maldacena.}    Suppose it were true that the extended action 
$\hat I(X,A,B)$ had a dual $\Z_N$ symmetry $B\to B+2\pi/N$.   This is not really the case, since that operation actually shifts the action by the topological invariant
$(2\pi\i/N)\int F/2\pi$, which is an integer multiple of $2\pi\i/N$ and not of $2\pi\i$.   Still, suspending disbelief, after orbifolding by $B\to B+2\pi/N$, the period of the field
$B$ would be reduced to $2\pi/N$.   To get back to a scalar field with the standard period $2\pi$, we could introduce $C=NB$.  In terms of $C$ the action
would be 
\be\label{toggox}\hat I_{1/N}(X,A,C)=\frac{\i}{N}\int_\Sigma \frac{CF}{2\pi}+I(X,A). \ee   This does not lead by any standard method to a well-defined quantum theory,
since the coefficient of the $CF$ term is not properly quantized.  Still, we see that,
 formally speaking, a $\Z_N$ orbifold of the dual variable would give the same thing we might hope to get by analytic continuation from $N$ to $1/N$.
As discussed in the introduction, replacing $N$ by $1/N$ is similar to considering an $N$-fold cover of the target space of the original theory, rather than an $N$-fold
quotient.   In this paper, we have found some evidence that, if $I(X)$ is a standard worldsheet action in string theory, then
 the theory associated to such an $N$-fold cover, if it exists, is a nonunitary logarithmic conformal field theory.

\section{An Alternative Formula For $G(z,\tau)$ }\label{pfunction}

The function $G(z,\tau)$ that we encountered in section \ref{partfn}  is a doubly-periodic meromorphic function, so, for fixed $\tau$,
it is can be uniquely characterized, up to a constant multiple, by its zeros and poles.
These  consist of  triple zeros at points of the lattice $\Z+\tau \Z$, and
simple poles at the non-zero half-lattice points.   

Let us compare this to the behavior of some doubly-periodic functions.
The Weierstrass $P$-function is defined by
\be\label{hudd}P(z,\tau)=\sum_{n,m\in \Z}\frac{1}{(z-n-m\tau)^2}. \ee
The sum requires some regularization, but does define a doubly-periodic function whose only poles are double poles at lattice points.   It is convenient to denote $P(z,\tau)$ as $x$.
The derivative of the $P$-function is
\be\label{udd} y=\frac{\partial P(z,\tau)}{\partial z} =-2\sum_{n,m\in \Z}\frac{1}{(z-n-m\tau)^3}. \ee
  Clearly $x$ is an even function of $z$ and $y$ is an odd function.  The definition of $y$ makes it obvious that $y$ has a triple pole at lattice points and no other poles; because $y$
  is an odd function of $z$, it vanishes at the nonzero half-lattice points $1/2, \tau/2, $ and $(1+\tau)/2$, as well as translates of these points by lattice vectors.  Since $y$ has only
  three poles (or rather one triple pole) in each fundamental cell of the lattice, it can have no additional zeros.
     Thus the function $1/y$ has the same zeros and poles as $G(q,w)$
  and for fixed $\tau$, the two must coincide, up to a constant multiple.    To fix the multiple, we just observe that near $z=0$,
  $1/y\sim -z^3/2$, while $G(z,\tau)\sim (2\pi\i z)^4/(4\pi\i z)=-\i(2\pi)^3 z/2$.    So
  \be\label{pludd} G(z,\tau)= -\i(2\pi)^3\frac{1}{y}.  \ee
  
  A standard result is that the  functions $x=P(z)$ and $y=P'(z)$ are related by
  \be\label{munx} y^2=4x^3-g_2x-g_3,\ee
  which is the usual description of a Riemann surface of genus 1 as an algebraic curve.
  Here $g_2$ and $g_3$ depend on $\tau$ only, and are modular forms of weights 4 and 6, respectively\footnote{$\Z^2_{\not=0}$ is the set of pairs of integers,
  not both zero.}:
  \be\label{unx}g_2(\tau)=\frac{60}{\pi^4}\sum_{m,n\in \Z^2_{\not=0}} \frac{1}{(m+n\tau)^4}, ~~g_3(\tau)=\frac{140}{\pi^6}\sum_{m,n\in\Z^2_{\not=0}} \frac{1}{(m+n\tau)^6}.\ee
    Eqn. (\ref{munx}) can be proved by comparing the expansions of the left and right hand sides near $z=0$.
  The statement $y=P'=\d x/\d z$ is equivalent to
  \be\label{nxix}{\d z}=\frac{\d x}{y}.\ee

\end{appendix}

\end{document}